# Internet Radio: A New Engine for Content Diversity?

Benjamin Compaine and Emma Smith
MIT Program on Internet & Telecoms Convergence
For the Telecommunications Policy Research Conference

October 2001

## INTRODUCTION

Throughout the world radio broadcasters are governed by policies and regulations promulgated by national legislatures and regulatory authorities. By 2001 radio programs were being delivered over the Internet. Traditional broadcasters could stream their content over this new channel, and brand new broadcasters could reach online audiences generally without requiring any governmental approval. Thus, as millions of households gain Internet connections, stakeholders such as the incumbent private and public broadcasters are faced with potentially new competitors for audience share and, in some cases, advertiser support.

In the United States much of the regulatory agenda has been rooted in the commitment of the original regulatory law enacted by Congress in 1934 and interpreted through the years by the Federal Communications Commission (FCC) to promoting program diversity in broadcasting. This goal of diversity has been accomplished in some nations by the programming efforts of government-controlled broadcasters. In the United States diversity was largely attempted by strictly limiting the number of licenses that could be controlled by an entity and by requirements for "public service" programming by all broadcasters.

Today, the larger U.S. broadcast groups see no reason to end the easing of ownership rules that followed the Telecommunications Act of 1996. Before the FCC they argued that "the Internet is one of the fastest growing media ever…. [D]ue to technological advances, there exists today more aveneues for the average citizen to receive information than there ever were before" [1]

The hypothesis of this study is that Internet radio has added diversity to the traditional over-the-air broadcast structure. To that end, this paper measures the level of diversity being created by Internet radio broadcasters (i.e. enterprises delivering entertainment and/or news and information content as an audio stream via the Internet). Stakeholders on both sides of this debate have asserted that the rise of Internet broadcasting has the potential to provide audiences with increased access to diverse programming; however, these assertions have been primarily based on anecdotal evidence systematically gathered data.

---

[1] Reply Comments of the National Association of Broadcasters. "1998 Biennial Regulatory Review – Review of the Commissions Broadcast Ownership Rules," MM Docket No. 98-35, August 21, 1998, pp. 5-6.

This study gathers and analyzes empirical evidence that provides support for the point of view that Internet radio is adding substantial diversity to the radio broadcasting industry. Diversity here is characterized as the vareity for program formats and the number of radio stations available to listeners under unrelated ownership. Uimately it finds that, by delivering diverse programming to a significant portion of the market, Internet radio broadcasters complement traditional radio and provide more overall diversity to audiences. If this the case, and assuming technological and industry developments make Internet radio programming available to a significant segment of radio audiences, regulators should consider the Internet in their analysis of the structure of the radio broadcasting industry. If the Internet is adding significant diversity to the radio broadcasting universe, regulators could further relax the ownership rules that currently govern the radio industry in many countries.

**Internet Radio Broadcasting**

That radio itself is a popular medium is underscored by its usage. At an estimated 967 hours per year per person, it is second only to the 1580 hours per person spent with teleivsion and greater than six times the hours spent reading newspapers.[2]

For the purposes of this study, Internet radio broadcasters are defined as entities that deliver entertainment and/or news and information content as an audio stream via the Internet. These audio streams may be delivered live or archived to be accessed on demand, but in both cases the audio files were initially created as programming to be delivered to an audience of more than one.

This definition means that the downloading of individual music files using services such as Napster is not being considered in this paper. It is a distinction intended to recognize the difference between programmed radio and "static" audio and centers on the issue of control. Services that allow users to program their own play lists (i.e. juke box style services that allow users to select songs and then have them assembled in the chosen order) are not considered radio in this instance.

---

[2] TV Basics: Consumer Media Usage, Television Bureau of Advertising, TVB Online. Data from Veronis, Shuler and Associates.

## DIVERSITY: A COMMON GOAL

In the United States, the Federal Communications Commission has had central to its long standing policy the furthering of diversity – never preceisely defined --  in broadcasting. In 1945, the Supreme Court counseled that the First Amendment "rests on the assumption that the widest possible dissemination of information from diverse and antagonistic sources is essential to the welfare of the public."[3]  More recently, former FCC Chairman William E. Kennard reaffirmed, "Broadcast remains the way that most Americans get vital information about their local community … (and so) retaining diversity of ownership of broadcast outlets is … vital to the democratic process."[4]

The two principles of localism and diversity underlying the FCC's efforts to regulate radio broadcasters stem from its belief that diversity is a commonly desired objective shared, or at least cited by, almost all players in the industry.  The principles that support the need for ownership regulations have been plainly articulated:

> First, in a system of broadcasting based upon free competition, it is more reasonable to assume that stations owned by different people will compete with each other, for the same audience and advertisers, than stations under the control of a single person or group.  Second, the greater the diversity of ownership in a particular area, the less chance there is a single person or group can have an inordinate effect, in a political, editorial, or similar programming sense, on public opinion.[5]

Since the beginnings of radio broadcast regulation, in the U.S. as elsewhere, the interest in promoting diversity has guided regulators and courts which have struggled to establish fair tactics but remained unanimous about the goal.

---

[3] <u>Associated Press v. United States</u>, 326 U.S. 1, 20 (1945)
[4] Press statement of FCC Chairman William E. Kennard regarding launch of biennial review of broadcast ownership rules.  March 12, 1998.
[5] 1998 Biennial Regulatory Review – Review of the Commission's Broadcast Ownership Rules and Other Rules Adopted Pursuant to Section 202 of the Telecommunications Act of 1996.  MM Docket No. 98-35.  Section B. p39.

# THE INTERNET

The Internet has already had a profound effect on radio broadcasting. Radio stations are learning to re-broadcast online, new dotcoms are taking advantage of the Internet as an accessible and regulation-free broadcast environment, while players and stakeholders are putting forth arguments about whether or not the Internet should be considered in today's discussions regarding the regulation of radio.

**Internet and Radio Broadcasting: The Situation Today**

In 2001 nearly 86% of the 12,500 radio stations in the United States had an Internet Web site. One fourth of all stations – 3093 – were available in real time via the Internet.[6] This was a nearly three fold increase from 1999.

While traditional radio stations were slow to start delivering audio programming over the Internet, new companies were springing up with radio-like formats. In 2001, online radio broadcasts were being delivered by providers as varied as non-media businesses and newspapers to pure Internet broadcasters ranging from NetRadio (www.netradio.com) to home-based operations such as Neurofunk (www.neurofunk.com).

Even at this nascent stage of Internet penetration there is a sizeable market for Internet radio programming. Audiences are apparently prepared to use the Internet to listen to the radio. By the start of 2001 7.3% of all Americans (or approximately 19 million people) were listening to radio programming over the Internet.[7] The number could be expected to rise dramatically in subsequent years. In one notable move, America Online, the largest Internet Service Provider with 31 million subscribers,[8] introduced in late 2001 a service called Radio@AOL.com with 75 "channels" of streaming programming.[9]

**The Internet and Regulation**

One proposed justification for the argument that government regulation of radio broadcasters should be relaxed is that the Internet, with its thousands of news, information and entertainment sites, provides significant and diverse programming to global audiences.

---

[6] Radio Advertising Bureau, *Radio Marketing Guide and Fact Book, 2001-2002*. New York, 2001, 36.

[7] The Arbitron/Edision Media Research Study VI "Steaming at a Crossroads" February 2001.

[8] "Worldwide AOL Membership Surpasses 31 Million," Press Release, America Online, Inc., Sept 11, 2001 at http://media.aoltimewarner.com/media/cb_press_view.cfm?release_num=55251177

[9] "AOL Music Announces New Initiatives In Online Music," Press Release, America Online, Inc. July 23, 2001 at http://media.aoltimewarner.com/media/cb_press_view.cfm?release_num=55252079

In order to understand the validity of arguments about the Internet's role in this debate, it is necessary to assess whether or not Internet broadcasters are, in fact, delivering diversity in ownership and format to the public.

### Federal Communications Commission (FCC)

Few government authorities officially take the Internet into account as an alternative source of radio programming. However, it is increasingly a topic of debate. The former chairman of the U.S. FCC assured stakeholders that new information channels had been considered during the FCC deliberations. He commented, "Although new technologies like the Internet and satellite delivery may be fundamentally changing the communications landscape, they do not yet command the time and attention of most consumers."[10]

Another FCC Commissioner, however, used the increasing prominence of the Internet, cable and other technologies to justify his assertion that ownership regulations are no longer needed to ensure diversity. He pointed out that "today broadcasters face such a fierce array of competitors – from cable operators, … internet service providers, wireless video systems, and direct satellite systems – that their previously supposed ability to influence the content and flow of information is greatly diffused. In sum, over time, as alternative means of communication … have proliferated in the marketplace, the burdens imposed on broadcasters by these restrictions have increased dramatically relative to the benefits that they produce."[11]

### Broadcasters. Many incumbent broadcasters believe that the Internet does increase the programming options available to their audiences and are using its growing popularity to support their effort to loosen regulatory restrictions on ownership limits and programming. When looking at diversity in viewpoints, the pitosion of many U.S.

---

[10] Separate Statement of Chairman William E. Kennard, In the Matter of the 1998 Biennial Regulatory Review. June 20, 2000.
[11] Joint Statement of Commissioners Powell and Furchtgott-Roth, In re Personal Attach and Political Editorial Rules, FCC Gen. Docket No. 83-484, at 5 and n.15.

broadcasters is that the FCC should "look at all media, including television, radio, cable, DBS, the Internet and newspapers….."[12]

## TECHNOLOGY EVOLUTIONS

The two technological evolutions that can be expected to play a particularly significant role in the advancement of Internet radio are mobility and broadband Internet access.

### Mobility

Mobility is a central component of radio. People listen to radio precisely because it's a medium they can employ while continuing on with other activities (i.e. driving, working) and it this same trait that attracts advertisers hoping to reach listeners outside of their homes. The proliferation of mobile Internet access has the potential to broaden the base of Internet radio. Emerging technologies are enabling new broadcasters to deliver interactive radio content to mobile listeners, and this mobility is already creating new opportunities for Internet radio broadcasters to compete with traditional AM and FM stations. Mobile Internet access has the potential to create a world in which consumers can access Internet radio stations without being connected to a computer and are able to listen to online broadcasters anytime and anywhere.

#### *New Mobile Alternatives*

There are several technologies emerging in the race to deliver radio programming to mobile devices and vehicles. One of the most hyped is satellite radio. In 1997 the FCC granted broadcast licenses to XM Radio and Sirius Radio which began delivering radio programming to cars in 2000 via satellite networks. Cellular phones, personal digital assistants and even video game consoles may increasingly be wireless Web-connected, which means that accessing the Internet will no longer be just a PC-based activity.

### Broadband

The second technological evolution that can be expected to expand Internet radio broadcasting is broadband connectivity. In the U.S., of the 56% of homes with Internet access, approximately 16% had broadband connectivity at the end of 2000. Predictions are that this may reach 46% by 2003.[13]  This proliferation is significant because it changes the way consumers use the Internet and also because it makes audio and video more accessible online.

---

[12] 1998 Biennial Regulatory Review – Review of the Commission's Broadcast Ownership Rules and Other Rules Adopted Pursuant to Section 202 of the Telecommunications Act of 1996. MM Docket No. 98-35. Section B, p.4.

***Media Usage.*** Media consumption is affected by Internet use, particularly in broadband households.  In the households of early broadband adopters, the Internet's share of media time surges to 21%, from 11% average for all households.[14]  Broadband also changes the geographical boundaries of media usage.  Broadband users are far more likely to search out and use audio and video content from around the world.  The "sheltered garden" of a local broadcast market could be transformed in a broadband world.[15]  This means that broadband Internet users are increasingly likely to use the Internet as another source of international news and entertainment programming.

## METHODOLOGY

Internet radio broadcasting is clearly a very young market.  The limited availability of data regarding Internet radio broadcasting means that, to date, only a small amount of analysis has been conducted in this area.

The paper draws on data from Arbitron, MeasureCast and Realguide.com, each of which will be discussed in more detail below, as well as from the individual radio station Web sites.  Arbitron was used as a source of data about the Internet radio stations garnering the greatest listenership in the United States. Data from MeasureCast was extracted and used as a benchmark to test the validity of the Arbitron numbers. Given that Arbiton and MeasureCast only provide data on the most listened to of their own primarily North American subscribers, a third data source, Realguide.com, was used as a source of information about the broader range of global radio broadcasting available on the Internet.

Some additional data was gathered from the individual station Web sites.  Data on Internet radio station listenership, format and technology platform was gathered from Arbitron's September 2000 ranking.  Each radio station site ranked by Arbitron was then visited online.  Data about each station's target market and programming language was added to the Arbitron data, as well as an assessment of whether the station was also broadcasting via an AM or FM channel. Because the Arbitron and MeasureCast rankings measure only the 75 most listened to Internet radio stations, detailed data was also gathered from Realguide.com in an attempt to understand the levels of format and language diversity available from the stations not securing highest Aggregate Tuning Hours (ATH).

---

[13] eMarketer: Security risks lie beneath broadband hype.  December 13, 2000.  Percentage of online households with broadband connectivity.

[14] *The Broadband Revolution: How Superfast Internet Access Changes Media Habits in American Households.* Arbitron/Coleman. 2000. New York. Page 3.  http://www.arbitron.com/radio_stations/home.htm

[15] *Ibid,.* p 19.

## Data Sources

### Arbitron and Server Log File Analysis

Arbitron measures radio audiences in local markets across the United States and has recently begun providing Internet information services for the advertising and commerce-supported Webcasting and online media markets as well as the advertisers and agencies that support it.[16] The data in this study was gathered by Arbitron in September 2000. The results were compared to the August 2000 results, which highlighted the way in which the results can shift from month-to-month as new subscribers join Arbitron and have their radio streaming measured.

Because Arbitron only measures the streaming conducted by their subscribers, it is possible for a "station" that wasn't ranked at all in one month to receive a high ranking the following month. This is simply because their log files weren't submitted in the first month and so weren't considered. For example, CFNY-FM held the number 13 spot in Arbitron's September, 2000 rankings. CFNY-FM didn't appear at all in the August 2000 results because they were not subscribers to the Arbitron service and so were not measured. This highlights an inherent problem with the Arbitron rankings: namely that, even if there are Internet radio stations receiving a high enough listenership to place in their top 75, they would not appear in the data if they are not Arbitron subscribers.

Arbitron ranks Internet audio providers according to a metric they call Aggregate Tuning Hours (ATH). ATH is based on a server-side measurement that captures all tuning to participating streamed media channels by compiling what Arbitron calls a near census of Internet tuning sessions.

Server Log File Analysis is one of the most common methods for measuring Web site traffic. Every time an event occurs on a server (e.g. a request is made or granted) the server writes a record of the event in a "log file." These log files can be analyzed to produce reports on the activity that occurred on that server during a specified time period.[17] For example, a log file that reads :

> 192.168.1.55 - - [14/Jun/2000:13:48:10 -0700] "GET encoder/live05.rm RTSP/1.0" 200 146835 [WinNT_4.0_6.0.6.94_play32_RN6C_en-US_686][d928cb60-3694-11d4-9071-0001023f3be2]

---

[16] Arbitron Internet Information Services corporate Web site. www.internet.arbitron.com

This says that a user running the WinNT 4.0 operating system at IP address 192.168.1.55 successfully (code 200) requested the file encoder/live05.rm with 146835 bytes on June 14, 2000 at 1:48 and 10 seconds PM PST using the RTSP/1.0 protocol. The user had the unique identifier [d928cb60-3694-11d4-9071 0001023f3be2] turned on in their media player. [18]

Server log file analysis does provide analysts with accurate stream counts; however it is not able to generate the demographic information about individual users that many advertisers are interested in. Also, because it is possible to alter log files relatively easily, this measurement technique does make it possible for subscribers to "cheat" the system and bolster their server ratings. A third issue associated with server log file analysis is the likelihood that Cumulative (i.e. unique user) numbers may be skewed due to the use of dynamic IP addresses or people sharing computers. This analysis does not provide an absolute measurement of individual users, but rather a measurement of individual IP addresses accessing the server.

### *MeasureCast and Active Event Monitoring*

MeasureCast provides Internet broadcasters, advertisers and media buyers with demographic information, as well as statistical analysis, regarding their Internet radio broadcasts. MeasureCast measures fewer sites than Arbitron but generates more in-depth information for their customers.

MeasureCast's methodology, called Active Event Monitoring, combines server log file analysis with a second technique called panel survey analysis, intended to add value-added demographic information to the data. Panelists share demographic information about themselves, participate in a series of exercises, with this data extrapolated to the demographics of the total audience.

The Active Event Monitoring system requires those broadcasters who want to receive measurement data to install a plug-in application to their streaming server that runs in the background of the server's regular functions. This plug-in "records data about each listening event from the server's broadcasts, and transmits that data using an encrypted channel to MeasureCast's centralized database server, where it is processed and stored. This data is then combined with demographic information from a statistically valid panel, representative of the known universe of streaming media users."[19]

---

[17] "An Analysis of Streaming Audience Measurement Methods," MeasureCast, Inc. Aug. 14, 2000.
[18] Ibid.

While the Active Event Monitoring system may, in the future, provide valuable demographic information that could be used to further understand the Internet radio audience, at this time the data most valuable is their measurement of Total Time Spent Listening (TSL) and Cume.  TSL is the total number of hours streamed by the broadcaster in the reported time period, and is the sum of the length of all listening events in that time period.  Cume makes an assessment of the actual number of unique individuals who accessed a broadcaster's radio streams.

MeasureCast's data is too limited to be considered as a complete data source for this study as the company only measures those stations that use RealNetworks and subscribe to their service.  Their development of a more in-depth measurement technique, however, does make their results useful as a point of comparison for the Arbitron results.

**Table 1.  Comparison of MeasureCast and Arbitron Rankings of those Stations Using the Real Networks Technology Platform**

| MeasureCast Top 10 | MeasureCast Ranking | Arbitron Top 10 Stations using Real Networks Platform | Arbitron Ranking |
|---|---|---|---|
| MediaAmazing | 1 | WABC-AM | 10 |
| WABC-AM | 2 | WPLJ-FM | 14 |
| WPLJ-FM | 3 | Tom Joyner Morning Show | 19 |
| Radio Margaritaville | 4 | Radio Margaritaville | 23 |
| KSFO-AM | 5 | KQRS-FM | 27 |
| KQRS-FM | 6 | WRQX-FM | 37 |
| WLS-AM | 7 | KLOS-FM | 39 |
| The Beat LA | 8 | WJZW-FM | 40 |
| Hard Radio | 9 | WLS-AM | 43 |
| WBAP-AM | 10 | WBAP-AM | 45 |

Source:  Arbitron and MeasureCast, 2000.

Table 1 compares a list of only those Arbitron-measured companies who also use the Real Networks platform to the relative position to the same stations ranked by MeasureCast.

---

[19] Ibid.

Table 1 clearly illustrates the limitations of relying on the MeasureCast rankings as the sole means of gauging Internet radio station usage. The top nine stations according to Arbitron do not appear at all in the MeasureCast rankings because they are not MeasureCast subscribers. The MeasureCast data is, however, still valuable as a point of comparison. As illustrated in Table 1, there is a relationship between the results provided by MeasureCast and those provided by Arbitron. Six of the 10 stations in MeasureCast's ranking also appear in Arbitron's top 10 when only the Real Networks stations are examined. KSFO-AM, ranked number 5 by MeasureCast, is 11[th] in Arbitron's listing. The few remaining discrepancies can be accounted for by the fact that they are stations not measured by Arbitron. Within this limited data, the similarity between the MeasureCast and Arbitron results are quite similar does lend support to the validity of the Arbitron data.

### Realguide.com

RealGuide is an online service provided by RealNetworks that provides searchable access to approximately 2500 [20] Internet radio broadcasters. While Arbitron and MeasureCast provide valuable data regarding some of the most popular Internet radio stations in the U.S., Realguide.com provides access to a broader listing of radio stations currently available online. Functioning primarily as an aggregator of Internet radio stations, Realguide.com provides direct links to station sites and allows prospective listeners to search the stations by language, format and location.

The stations listed by Realguide.com are important even if they individually garner only small audiences. This is because every one of the 2500 stations accessible from Realguide.com, as well as all the other small Internet radio stations available online, do provide consumers with access to new radio owners and programming. The 2500 stations listed at Realguide.com represent a broader range of what is available than is measured by Arbitron.

### Measurement Bias

Note that the data provided by Arbitron and MeasureCast is not complete and that, because each only measures the audio streaming being conducted by their own subscribers, the results have inherent limitations. Arbitron has approximately 900 subscribers, including radio broadcasters and service providers. Each of these subscribers pays a fee so that Arbitron will track their audio streaming. Arbitron generates a monthly report that ranks their subscribers and provides data on the top 75 performers. It is the data on these top 75 performers, each one an Arbitron subscriber, that is available for analysis.

---

[20] The actual number of stations varies slightly as broadcasters join and leave Realguide.com. Also, there may be some stations that do not fit into each of the categories chosen for this research.

MeasureCast subscribers, with the exception of a few independently streamed stations, all stream through Real Broadcast Networks. This means that many large stations being streamed by Akamai, iBEAM Broadcasting or Activate, for instance, are not considered in their rankings.

## Data Categories

In order to provide meaningful and replicable results, it was important to find accessible data points that could provide valid measurements of the diversity in programming being delivered by Internet audio providers.

For this study, the data categories gathered from Arbitron, MeasureCast and the Web are:

- Format
- Internet Only vs. Traditional broadcaster
- Owner
- Location of Primary Target Market
- Language

***Format***.. Format diversity, in this paper, is considered to be the variety of radio program types being delivered by online radio broadcasters. Format definitions are taken from the standard Arbitron guide used to delineate types of radio programming. By examining the range of formats being delivered online it may be possible to provide a sense of the programming diversity available online. Given that much of the concern around diversity is focused on news and information, particular emphasis is on the proportion of Internet radio broadcasting is news and information compared to music or other entertainment and other entertainment programming.

***Internet Only vs. Traditional broadcaster***. There are two kinds of Internet radio broadcasters: those who create programming solely for distribution over the Internet, and those who already distribute programming via traditional broadcast spectrum and are also delivering that content online. This distinction is important because it speaks to the ability of new broadcasters to compete with large incumbent broadcasters by taking advantage of the Internet as a lower cost broadcast process.

***Owner.***  Diversity in source, or enterprise affiliation, has long been an important measurement for regulators attempting to measure market power.  It is generally held – through not empirically proven -- that fewer owners will lead to less diversity in programming. "Source diversity" is measured by  the number of owners.  This criterion is consistent as defined by the FCC.

***Location of Target Market.*** One of the new categories available for study with the arrival of Internet radio is the location of the target market.  Anyone with Internet access can listen to any Internet radio station,  so broadcasters are not limited in reach to consumers in the footprint of their signal's footprint.

While an Internet broadcaster can, at least in theory, reach listeners anywhere in the world, many still focus their online efforts on a local audience.  This is significant because it speaks to the diversity of programming being offered online.  The intent of the broadcaster, in terms of the audience it believes it is reaching, will affect the content being provided (i.e. New York weather vs. Berlin weather), and so the objective of this data category is to quantify the geographical focus on the Internet radio programming being broadcast online.

***Language.*** Another data category that can provide interesting insights into the diversity offered by Internet radio broadcasters is language of the content.

**Market Share**

For the purposes of this study, Internet audio broadcasters delivering content online are all considered to be in the same geographic market because listeners anywhere can access them.  Traditionally the market is defined according to Arbitron's geographic regions that, essentially, determine that a broadcaster is in the same market as another radio broadcaster if the same listener can access both sources.  In other words, there are geographic boundaries.  Online, the question of geography is different, because people are not limited to accessing online broadcasters located in their area.  Broadly speaking, anyone delivering online radio programming over the Internet is therefore considered to be in the same geographic market, competing for the same audiences.

That said, gathering data about the location of the primary target audience for each of these Internet broadcasters should offer some indication of whether or not the majority of Internet radio programming is being accessed nationally, internationally, or primarily in a local market (i.e. a metropolitan area).

<div align="center">**FINDINGS**</div>

**Format**

There are both similarities and substantial differences in the types of content provided by Internet radio, as seen in Table 2. The number of stations featuring news and talk formats predominates regardless of the universe being measured: the 75 most popular Internet sites, the larger aggregation of RealGuide sites, or the census of all US licensed broadcasters. But beyond that variances emerge. Perhaps the most noticeable is the category called World Music, which accounts for 8.2% of the 2500 global sites covered by RealGuide. This category is not recognized by Arbitron nor the U.S. industry overall. Contemporary Hits are the second most popular format among the Arbitron Internet sites, but less than half the ratio in RealGuide and account for just over 3% of US broadcasters. On the other hand, Country Music, the second most common format of U.S. radio stations, fades substantially when incorporating the far more international group of RealGuide stations.

Taken together, this comparison suggests that the variety of formats available online is different from and far more varied than what is available via conventional broadcasting.

**Table 2: Top Internet Formats Compared to U.S. Broadcast Formats, 2000**

|                    | Arbitron 75 | RealGuide | All U.S. AM/FM |
|--------------------|-------------|-----------|----------------|
| News/talk          | 18.7%       | 17.4%     | 16.0%          |
| Contemporary Hits  | 10.7        | 4.5       | 3.2            |
| Country            | 8.0         | 3.6       | 14.7           |
| Classical          | 6.7         | 3.7       | 2.9            |
| Jazz               | 5.3         | 5.5       | 2.5            |
| World Music        | NA          | 8.2       | NA             |

NA: not applicable—this format is not recognized

Sources: Arbitron Webcast rating, August 2000; RealGuide, http://realguide.real.com/tunner; Broadcast and Cable Yearbook, 1999

**Internet Only vs. Traditional Broadcaster**

Are traditional broadcasters dominating the online broadcast world or are new Internet-only broadcasters securing meaningful market share? This question asks whether the Internet is opening the door for new voices hoping to reach radio listeners, or whether it is, in fact, simply serving as an additional conduit that traditional broadcasters can use to distribute content.

**Table 3.  Internet Only Broadcasters vs. Internet with AM/FM Affiliate**

| Type of Owner | # of Stations owned | % of stations owned | % of total ATH |
|---|---|---|---|
| Internet Only | 36 | 48.0 | 57.5% |
| Internet with AM/FM Affiliate | 39 | 52.0 | 42.5 |

Source:  Compiled from Arbitron Webcast Ratings, September, 2000.  Top 75 Aggregate Tuning Hours (ATH).  www.arbitron.com

As seen in Table 3, the 75 Internet radio broadcasters measured by Arbitron were almost equally divided between Internet-only and traditional broadcast broadcasters.  The picture changes somewhat when the two groups are measured based on Aggregate Tuning Hours (ATHATH), also tabulated in Table 3.  Internet-only radio stations account for 58% of the total hours spent listening to Internet radio, compared to 42% for traditional broadcasters. This suggests, that at this stage at least,  that, at this stage at least, the new Internet-only broadcasters may be more attuned to the needs of the Internet radio audience.

**Ownership**

There are 24 companies that own one or more of the top 75 radio stations ranked by Arbitron according to the Aggregate Tuning Hours measurement.  Of these, 14 have only a single property among the top 75. As seen in Table 4, the two largest, NetRadio and ABC Radio, account for 56% of the total number of stations.   NetRadio has almost twice as many properties as ABC Radio, with 27 stations.

The gap between NetRadio and ABC Radio is even greater when measured according to Aggregate Tuning Hours.  As seen in Table 4, NetRadio captures 43% of the listening hours, although it owns only 36% of the stations it captures 43% of the listening hours although it owns only 36% of the stations.  The ABC stations, on the other hand, capture only 16.9% of the ATH while they own 20% of the stations.

The top four Internet radio owners in Table 4 together own 64% of the top 75 Internet stations which combine to account for 67.1% of the total ATH.  Studies of licensed radio broadcasters, however, have found a higher ownership concentration

Table 4.  Ownership Diversity Among Arbitron's Top 75 Internet Radio Stations

| Owner | Internet Only or Broadcaster | # of Stations owned | % of top 75 stations owned | % ATH owned |
|---|---|---|---|---|
| **NetRadio** | **I** | **27** | **36.0** | **43.3** |
| ABC Radio | B | 15 | 20.0 | 16.9 |
| New Wave Broadcasting LP | B | 3 | 4.0 | 3.7 |
| Fisher Broadcasting | B | 3 | 4.0 | 3.2 |
| Bonneville International | B | 3 | 4.0 | 2.7 |
| CHUM Group | B | 2 | 2.7 | 1.2 |
| Citadel Communications | B | 2 | 2.7 | 2.0 |
| Corus Entertainment | B | 2 | 2.7 | 1.8 |
| **Enigma Digital** | **I** | **2** | **2.7** | **5.7** |
| Inner City Broadcasting Corp. | B | 2 | 2.7 | 1.7 |
| EXCL Communications | B | 1 | 1.3 | 0.7 |
| **eYada** | **I** | **1** | **1.3** | **1.2** |
| **Global Media** | **I** | **1** | **1.3** | **0.5** |
| Ingleside Radio, Inc. | B | 1 | 1.3 | 0.6 |
| One-On-One Sports | B | 1 | 1.3 | 1.3 |
| Pacific Lutheran University | B | 1 | 1.3 | 1.8 |
| **Radio Margaritaville LLC** | **I** | **1** | **1.3** | **1.4** |
| **Salem Comm. Corp** | **I** | **1** | **1.3** | **1.3** |
| Santa Monica College | B | 1 | 1.3 | 1.2 |
| Scottish Media Group | B | 1 | 1.3 | 2.8 |
| Shaw Communications | B | 1 | 1.3 | 2.1 |
| Sunburst Media L.P. | B | 1 | 1.3 | 1.5 |
| **The Broadcastweb Network, Inc.** | **I** | **1** | **1.3** | **0.8** |
| Texas Country Connection | B | 1 | 1.3 | 0.6 |

Source:  Arbitron Webcast Ratings, September 2000.  Top 75 Aggregate Tuning Hours (ATH) www.arbitron.com. Internet-only owners shown in boldface.

in local markets.  In 2000 the FCC found that the top four radio owners generally accounted for about 12% of the total number of stations[21], though as much as 90% of advertising revenue in some local markets.[22]

Table 4 further shows that 25% of the companies represented own Internet-only stations.  Primarily because of NetRadio's large share of the top ranked stations and the total ATH, Internet only broadcast owners represent only a quarter of the owners of

[21] FCC.  Review of the Radio Industry, 2000, July 2000, calculated from Appendix A and Appendix B.
[22] Ibid,  Appendix D.

Internet radio stations, yet they are account for 54% of the Internet radio listening online. Thus, Internet-only radio station owners, according to this preliminary data, have been able to successfully compete with traditional broadcasters for a substantial share of the online audience.

**Table 5.  Internet Radio Stations by Location of Target Market**

| Location | # of stations with target market in this location | % of top 75 stations with target market in this location | % ATH delivered to target market in this location |
|---|---|---|---|
| National | 35 | 46.7 | 55.7 |
| San Francisco/California | 8 | 10.7 | 8.4 |
| Washington, DC | 6 | 8.0 | 5.4 |
| Dallas/Texas | 3 | 4.0 | 3.0 |
| Seattle | 3 | 4.0 | 3.2 |
| New York | 3 | 4.0 | 5.6 |
| Calgary | 2 | 2.7 | 1.8 |
| Chicago | 2 | 2.7 | 1.6 |
| Detroit/Windsor, ON | 2 | 2.7 | 1.2 |
| Los Angeles | 2 | 2.7 | 1.7 |
| Minnesota | 2 | 2.7 | 1.9 |
| London | 1 | 1.3 | 2.8 |
| Toronto | 1 | 1.3 | 2.1 |
| Western Washington | 1 | 1.3 | 1.8 |
| New England | 1 | 1.3 | 1.4 |
| Boston, Chicago, LA, New York * | 1 | 1.3 | 1.3 |
| Columbus, OH | 1 | 1.3 | 0.6 |
| Portland, ME | 1 | 1.3 | 0.6 |

*\* One on One Sports (www.1on1sports.com) is a single Web site that provides access to four separate radio station sites.  Each station can be accessed from the One on One Sports portal but each has an affiliate in a different city.*

Source:  Arbitron Webcast Ratings, September, 2000.  Top 75 Aggregate Tuning Hours (ATH) www.arbitron.com

**Location of Target Market**

Traditional AM and FM broadcasters are limited in  where their radio audiences can be located.  Online, however, it is possible for any Internet radio listener to access a radio station from anywhere in the world.  While data is not available to measure the location of the listeners to the top 75 radio stations measured by Arbitron, a more useful tool for analyzing the content diversity is the location of the station's target market.  If, for instance, a  third of the top 75 stations were gearing their broadcasts for a New York audience by delivering news, weather and traffic reports specific to New York, this would suggest that diversity in programming is being limited by the location of the top broadcasters, even though their content is available internationally.

As seen in Table 5, 47% of the 75 Internet radio stations measured by Arbitron are created for a national market, while the balance are targeted to listeners in specific locations.  By comparison, when the target market locations are compared according to listening hours, 56% of the actual tuning hours are being spent listening to the national stations.  Thus the national, Internet-only stations are attracting a proportionately greater share of listening time than the local, licensed stations that are also broadcasting online. This makes sense given that the AM and FM stations have their traditional over-the-air signal by which listeners can access their programming.  All listening to the Internet-only stations must take place online.

Of those who choose to listen to the local AM and FM radio stations via online, early research from Edison Media finds that 56% choose stations that are from their market, compared to 34% who choose stations in other markets, and 6% who choose stations from other countries.[23]  This finding was substantially replicated in a study from Arbitron/Coleman Research that measured use of Internet radio stations by users with dial-up connections.[24]  This study noted an apparent red flag for local broadcasters, however.  It found early indications that the popularity of local Internet radio stations might wane as Internet connection speeds increase.

Specifically, the Arbitron/Coleman study found that broadband users are more likely than dialup users to listen to out-of-market stations: 41% of broadband users, compared to 35% of dialup users, choose U.S. radio stations from outside their own market. Furthermore, 17% of broadband users, compared to 10% of dialup users, listen to radio stations streaming from other countries. [25]  There are a number of possible reasons for this phenomenon. First, it may simply be that people with higher speed connections tend to be more adventurous online. Their faster connections mean that they can use a wider range of all Internet services than users with slower connections. In the absence of further research, it is also possible to speculate that as broadband Internet access was initially more readily available to urban dwellers, they may be more likely to listen to Internet radio from other markets regardless of their connection speed. They may be a demographic group that also has a higher than average income or education level, or who have traveled more than their counterparts with slower connection times, or may be more heavily weighted to mobile professionals or to recent immigrants, each of which might suggest an increased likelihood to listen to radio from outside their own market, egardless of connection speeds.

---

[23] Edison Media.  "Internet Study V: Startling New Insights About the Internet and Streaming," New York, September 2000.
[24] Arbitron/Coleman,  "The Broadband Revolution: How Superfast Internet Access Changes Media Habits in American Households,"  New York, October 2000.
[25] Ibid.

**Language**

Only one of the Internet radio stations rated in the top 75 by Arbitron delivers programming in a language other than English. This is perhaps not surprising. It would be very difficult for non-English language broadcasters to garner the broad listenership required to compete with the English language country music and classic hits stations that secure Arbitron's top spots.

A far different picture emerges upon examination of the approximately 2500 Internet radio stations listed by Realguide.com. As seen in Table 6 there are a significant number of radio stations now available via the Internet delivering programming in languages other than English. Many of the Internet radio stations listed by Realguide.com are delivering programming in languages ranging from Thai to Lithuanian and Mandarin. While these stations may not attract the large American audiences of the Arbitron-ranked sites, the fact that they are available over the Internet does support the notion that the Internet is making available diverse programming in a way that traditional radio is not. Thus, Internet users anywhere in the world can access radio programs from almost any country and in many languages.

While it is difficult to gauge whether the delivery of one non-English language station over the Internet reflects demand, it is possible to compare this availability to the number of non-English language stations available from traditional broadcasters. Out of the 12,467 radio stations broadcasting in the United States, 642 deliver programming in languages other than English. This includes 528 Spanish stations; five Portuguese, four each French, Greek and Polish; three in a Chinese dialect; two Arabic; and one each in Russian, Eskimo, Vietnamese and Filipino.[26]

**Table 6.  Languages Available At Internet Radio Stations Listed By Realguide.Com**

| Language | # of stations | % of stations |
|----------|---------------|---------------|
| English | 1596 | 78.85 |
| Spanish | 82 | 4.05 |
| German | 63 | 3.11 |
| Portuguese | 55 | 2.72 |
| French | 41 | 2.03 |
| Italian | 19 | 0.94 |
| Dutch | 17 | 0.84 |
| Polish | 11 | 0.54 |
| Arabic | 11 | 0.54 |
| Greek | 10 | 0.49 |
| Icelandic | 9 | 0.44 |
| Russian | 9 | 0.44 |
| Croatian | 8 | 0.40 |
| Hindi | 8 | 0.40 |
| Swedish | 8 | 0.40 |
| Czech | 7 | 0.35 |
| Turkish | 7 | 0.35 |
| Cantonese | 6 | 0.30 |
| Slovenian | 6 | 0.30 |
| Thai | 5 | 0.25 |
| Estonian | 4 | 0.20 |
| Farsi | 4 | 0.20 |
| Hebrew | 4 | 0.20 |
| Galician | 3 | 0.15 |
| Latvian | 3 | 0.15 |
| Norwegian | 3 | 0.15 |
| Romanian | 3 | 0.15 |
| Hungarian | 2 | 0.10 |
| Japanese | 2 | 0.10 |
| Mandarin | 2 | 0.10 |
| Finnish | 2 | 0.10 |
| Punjabi | 2 | 0.10 |
| Serbo-Croat | 1 | 0.05 |
| Slovak | 1 | 0.05 |
| Bulgarian | 1 | 0.05 |
| Catalan | 1 | 0.05 |
| Danish | 1 | 0.05 |
| Korean | 1 | 0.05 |
| Lithuanian | 1 | 0.05 |
| Luxemborgeois | 1 | 0.05 |
| Swiss | 1 | 0.05 |
| Tunisian | 1 | 0.05 |
| Urdu | 1 | 0.05 |
| Vietnamese | 1 | 0.05 |
| Source:  RealGuide.  http://realguide.real.com/tuner | | |

Internet radio stations offer significantly more language diversity than traditional over-the-air broadcasters. For example, the five Portuguese language AM and FM stations in the U.S., are available only to the audiences within range of their signals.

On the other hand, each of the 55 Portuguese stations available at Realguide.com is available to listeners anywhere in the world. In other words, one Portuguese station online potentially provides language diversity to a greater number of listeners than all five of the offline stations.

Furthermore, whereas the 12,467 traditional radio broadcasters in the U.S. represent only 11 languages, there are Internet radio stations delivering programming in 44 different languages. Most of these broadcasts originate outside the United States; however, to the listener in search of diversity the origin is of little consequence. The lesson of language diversity is equally applicable to any national market

## ANALYSIS

An analysis of the most listened to Internet radio stations ranked by Arbitron as well as the more eclectic offerings listed by Realguide.com provides some useful insights into the level of diversity being delivered online. There are two overarching findings that are particularly relevant from a policy perspective.

- The availability of radio programming globally has been dramatically enhanced by the reach of the Internet. There are more diverse programming options available, especially for audiences in rural areas and small towns as well as smaller countries with only a few over-the-air signals available,. This programming diversity does come, in part, from the availability of Internet-only broadcasting newcomers. Even if those dotcom broadcasters were unable to compete effectively in the marketplace, however, it can still be argued that audiences would have access to more diverse programming by virtue of being able to use the Internet to listen to traditional stations from outside their own market.

---

[26] *Broadcasting and Cable Yearbook, 2000 (*New Providence, NJ: R.R. Bowker Co., 1999). U.S. Radio Formats by State and Possession. 1999.

- While there may be a degree of ownership concentration among Internet radio broadcasters, this study has found that the Internet has made it possible for new players to enter the radio broadcasting market. Whereas these players may not have had the ability to win spectrum licenses and the approval to broadcast via a traditional terrestrial channel, new owners like NetRadio have been able to gain significant market share online.

Other findings in this study:

**Format**

The Internet has added to the number of radio formats available. Niche broadcasters are delivering specialized formats such as Tech Talk and Soundtracks but, in addition, even slightly more traditional formats such as Bluegrass and World Music are made available to all Internet radio listeners rather than only to those listeners in the large centers likely to have access to such programming via terrestrial channels. It is, therefore, audiences in small markets and with niche interests that have the most to gain from the diverse formats available online.

**Internet Only vs. Traditional Broadcaster**

This research supports the notion that the Internet, with its lower barriers to entry and open regulatory environment, provides opportunities for new, start up stations to compete with established broadcasters.

It is worth noting that the most listened to Internet radio stations are owned by NetRadio, an Internet-only broadcaster with no AM or FM affiliate stations. While traditional AM and FM broadcasters currently own more stations than Internet-only broadcasters, they receive less than half of the listeners' tuning time. NetRadio alone receives 43% of the total ATH listeners spend with Internet radio stations.

**Ownership**

This analysis of Internet radio station owners does demonstrate a relative concentration among a small number of players. The top two owners (NetRadio and ABC Radio) deliver more than 50% of Internet radio listening time across the country. Given that a few owners account for a significant proportion of the national audience, it is foreseeable that they will therefore gain access to an equal proportion of advertising revenue. In short, a few large Internet radio stations are likely to account for a similar proportion of ad spending.

While there are early signs of ownership concentration, there also appears to be a steady influx of new players entering the radio broadcasting market. These new players, if they are able to survive, will represent an increase in the overall number of owners providing radio broadcasting alternatives to consumers.

Furthermore, because Internet radio stations are able to broadcast from international locations, the Internet has made it possible for stations to diversify the ownership pool without actually having a physical presence in a specific national market. International broadcasters, then, may eventually be able to compete in the U.S. market for listeners and perhaps even advertising dollars, while U.S.-based Internet radio owners can aim for audiences globally.

Of particular note among the new influx of broadcast owners is that the most listened to Internet radio broadcaster is NetRadio. This highlights the fact that Internet broadcasting has added to the overall diversity in ownership among radio broadcasters. NetRadio owns twice as many stations as its nearest competitor, a traditional media company.

**Location of Target Market**

All but one of the top-ranked Internet-only radio stations focus on national rather than local content. Most of the Internet radio broadcasters who also broadcast over the air, however, are simply re-transmitting their locally focused AM and FM based stations.

While 53% of the Internet radio stations ranked by Arbitron are targeted to listeners in a specific city, it is the nationally focused Internet radio stations that capture the majority of the listeners' attention. The Internet-only broadcasters account for 56% of the tuning hours devoted to Internet radio. This may suggest that the promise of the Internet is coming true and that it is the new, Internet-only competitors who are dominating online and providing new content alternatives for consumers. Conversely this may be a reflection of the different goals that Internet and traditional radio stations have for the delivery of content online. While Internet-only stations rely on the Internet for 100% of their listenership, traditional broadcasters may choose to focus on their local markets and only invest a small percentage of their budgets on the Internet, viewing the platform as a tool for gaining a marginal increase in listenership rather than as a core business.

**Language**

While the most listened-to Internet radio broadcasters in the United States offer very little non-English language programming, the more inclusive Realguide.com listing demonstrates that there are many non-English language stations streaming content over the Internet. Compared to the two languages (including English) being represented by the Arbitron ranked stations and the eleven available on the 12,467 AM and FM stations broadcasting in the U.S., there are 44 languages represented at Realguide.com. This means that consumers gain access to significantly more language diversity with the advent of Internet radio broadcasting, something that is particularly relevant for consumers living in small markets with fewer AM and FM radio stations.

## IMPLICATIONS

The results of this research suggest a range of implications for radio broadcasters, regulators and listeners. The competitive landscape is being affected by the arrival and increasing popularity of Internet radio. The ability of Internet broadcasters to sidestep traditional barriers to entry is highlighted by NetRadio's high online visibility, and issues surrounding diversity and local programming are certainly worthy of further investigation as Internet broadcasters increasingly deliver global programming to audiences around the world.

The Internet has already changed the radio broadcasting marketplace and introduced new sources of diversity to audiences. With rapid advances in mobile Internet access technology and the deployment of high-speed Internet connectivity in many developed markets, Internet radio usage should continue to grow. Audiences with broadband connections are better able to access audio files and, once it becomes possible to access Internet radio stations in the car and from other mobile devices, AM and FM radio broadcasters' traditional prominence as the mobile media source may be further threatened. On the other hand, the ability to reach global audiences and the continued surge in the use of Internet radio will also create new opportunities for traditional radio broadcasters to expand their offerings and attract new audiences.

It is important to note that the implications discussed here stem from the data available in today's young Internet radio marketplace. These findings are likely to be reinforced as technology evolutions such as mobility and broadband access expand the market and generate additional interest in Internet radio. The findings would also be reinforced if this research was expanded to include a consideration of audio download services such as Napster

## Competition

A clear implication of this early research into diversity in Internet radio broadcasting is that the Internet is presenting traditional radio broadcasters with a new source of competition. The Internet is allowing new competitors to enter the market and, at least so far, it is these new, Internet-only broadcasters who are capturing the majority of the listening time nationally.

What is less clear, at the time of this writing, is whether new, Internet-only broadcasters will compete effectively with terrestrial broadcasters for advertising revenue. So far Internet radio stations have not had a significant impact on ad spending. However as ad insertion technologies improve and Internet radio use increases, it is possible that Internet radio stations and terrestrial radio stations will find themselves competing for advertising revenue.

Other, hybrid forms of audio, are also developing. On-demand community radio—essentially radio-like formats that are archived for ither download or streaming play back – may form a bridge between real time radio and recorded music and talk.[27]

The success of new competitors in the radio broadcasting space implies that there is an argument to be made by radio station owners for the further relaxation of government ownership restrictions. There are early signs that the Internet is allowing for competition from a new group of broadcasters.

## Concentration

There are early hints of ownership concentration among Internet radio broadcasters. The net effect of the Internet to date, however, has been to expand the number of owners delivering radio programming to audiences.

Should levels of ownership concentration continue to increase, one possible implication is that this may lead to a similar concentration in advertising revenue. In AM and FM markets one measure of ownership concentration is the percentage of advertising revenue accruing to the largest broadcast owners. While share of advertising information is not yet available for Internet radio stations, percentage of ATH is likely to be a strong indicator of each owner's ability to capture a corresponding share of advertising revenue.

---

[27] Michael Brown, "Internet Radio On-Demand," *Feedback,* 42:3 Summer 2001, pp. 19-23.

**Barriers to Entry**

In the traditional world of radio broadcasting one of the major barriers to entry for prospective newcomers is access to spectrum. With a limited spectrum available, new broadcasters must apply for a license to that spectrum, often a long and costly process that serves as a significant hurdle for many would-be broadcasters. Whereas access to spectrum used to be the most significant barrier to entry for new radio broadcasters, this barrier is non-existent for Internet radio broadcasters.

Instead the key barrier to entry for Internet radio broadcasters is the more universal one of gaining access to capital and workable business model. The importance of access to capital as the primary barrier to entry implies that start up, niche stations must have solid business models and the ability to generate revenue and be profitable or be otherwise subsidized.

**Competitive Advantage**

NetRadio's early dominance begs the question of whether Internet radio broadcasters in fact have an advantage over traditional AM and FM stations who are often slower to make the move to the Internet. Internet-only radio broadcasters are not subject to ownership restrictions, do not need to pay licensing fees, and have more freedom about the programming content they delivery. Also, because the Internet is their primary delivery channel, Internet-only stations may be more focused and more technically proficient than their AM and FM counterparts.

On the other hand, the ability of Internet-only stations to compete in the radio broadcasting arena has implications for traditional broadcasters hoping to establish a strong competitive advantage. Whereas the dotcom broadcasters must start from scratch, AM and FM stations already have a team of reporters, editors and content creators developing programming 24 hours a day. Unlike their Internet-only counterparts, AM and FM broadcasters have strong brand identities and organizational infrastructures that they can leverage on the Web. Likewise, traditional broadcasters have established, real world marketing engines at their disposal. They can use their AM and FM channels to promote their Web broadcasts and launch cross-platform campaigns to benefit both their on and offline efforts.

**Local Programming**

One of the key implications to stem from this research, and a good candidate for further research, is the likelihood that Internet radio broadcasting may lead to a decrease in the availability of local programming such as news, local-issues talk shows, traffic and weather.  The majority of the content being broadcast by Internet radio stations is either non-geocentric in focus, or targeted at consumers in a few major centers. This means that, while traditional broadcasters are expected to deliver local content, Internet radio stations are free to create programming for people living in Atlanta, Paris, Melbourne and Cape Town.  To the extent that government regulators continue to expect localism from radio license holders in the United States, Internet-only radio broadcasters may not further that goal.

## CONCLUSIONS

The results, analysis and implications put forth in this study generally support the initial hypothesis that by the measures used Internet radio has added diversity to the traditional over-the-air broadcast structure.  That is, audiences have access to a greater diversity of formats, channels, owners, languages, geographically focused programming and content distributors than they would if Internet radio broadcasts were not available.

While these baseline results do suggest that Internet radio has added diversity to the marketplace, it is critical to remember that the Internet radio broadcasting industry is still very young, and that economic and industry forces may affect the ability of smaller, niche broadcasters to continue providing this diversity.  It is necessary and likely, therefore, that additional research be conducted to monitor this industry over time.  Issues including ownership concentration, the ability of distributors to create bottlenecks to hinder diversity, the link between broadband connectivity and the use of out-of-market radio stations, the cost of bandwidth and connectivity to the consumer and the revenue models that will drive Internet radio broadcasting are thought to be particularly worthy of further study.

Overall, however, Internet radio broadcasters have already added new station options to the marketplace and provided a new source of competition for traditional broadcasters.  This research depicts an increase in diversity as a result of Internet radio broadcasting.  The radio universe as a whole has already been expanded significantly and positively and, as the industry evolves, this expansion is likely to continue.

**Bibliography**

America Online, Inc. "Worldwide AOL Membership Surpasses 31 Million," Press Release. Sept 11, 2001. http://media.aoltimewarner.com/media/cb_press_view.cfm?release_num=55252177.
America Online, Inc "AOL Music Announces New Initiatives In Online Music," Press Release. July 23, 2001. http://media.aoltimewarner.com/media/cb_press_view.cfm?release_num=55252079.

Arbitron/Coleman  The Broadband Revolution: How Superfast Internet Access Changes Media Habits in American Households.  New York. October 2000. http://www.arbitron.com/radio_stations/home.htm

Arbitron/Edison Media Research.  Internet Study V: Startling New Insights "About the Internet and Streaming.  New York, September 2000.

Arbitron/Edison Media Research.  Internet Study VI: Internet Study VI: Streaming at a Crossroads.  New York, February, 2001.

Associated Press v. United States, 326 U.S. 1, 20 (1945)

Broadcasting and Cable Yearbook, 2000. New Providence, NJ: R.R. Bowker Co. 1999.

Brown, Michael. "Internet Radio On-Demand" Feedback. 42:3 Summer 200., pp. 19-23.

EMarketer.  Security risks lie beneath broadband hype.  December 13, 2000.

Federal Communications Commission.  Biennial Review Report.  Adopted May 26, 2000. July 2000.

Federal Communications Commission. 1998 Biennial Regulatory Review.  Review of the Commission's Broadcast Ownership Rules and Other Rules Adopted Pursuant to Section 202 of the Telecommunications Act of 1996.  MM Docket No. 98-35.  Section B.

Kennard, Chairman William E. Separate Statement in the Matter of the 1998 Biennial Regulatory Review.  June 20, 2000.

MeasureCast, Inc.  An Analysis of Streaming Audience Measurement Methods.  Aug. 14, 2000.

National Association of Broadcasters. Reply Comments before the Federal Communications Commission.  In the Matter of 1998 Biennial Regulatory Review.  MM Docket No. 98-35. Accssed at http://gullfoss2.fcc.gov/prod/ecfs/retrieve.cgi?native_or_pdf=pdf&id_document=2137250001

Powell, Commissioner and Commissioner Furchtgott-Roth.  Joint Statement.  FCC Gen. Docket No. 83-484.

Radio Marketing Guide and Fact Book for Advertisers, 2001 to 2002.  Radio Advertising Bureau Headquarters and National Marketing Department, New York, NY.  2001.

Telecommunications Act of 1996.  Title 2, Section 202.

Television Bureau of Advertising. "TV Basics: Consumer Media Usage," TVB Online. Data from Veronis, Shuler and Associates. Accessed Sept 20, 2001 at http://www.tvb.org/search/docs/build_frameset.cgi?url=/tvfacts/trends/tv/index.html&my_section=tvfacts